\documentclass[conference]{IEEEtran}

\usepackage[T1]{fontenc}
\usepackage{newtxtext}
\usepackage{newtxmath}
\usepackage{microtype}
\usepackage{graphicx}
\usepackage{booktabs}
\usepackage{array}
\usepackage{tabularx}
\usepackage{url}
\usepackage{cite}
\usepackage[hidelinks]{hyperref}
\usepackage{xcolor}
\usepackage{balance}

\hypersetup{
  colorlinks=false,
  pdfauthor={Jepson Taylor, Chris Brousseau, Jordan Hildebrandt, Kelli Quinn},
  pdftitle={ZitPit: Consumer-Side Admission Control for Agentic Software Intake}
}

\definecolor{boxfill}{RGB}{247,248,250}
\definecolor{boxstroke}{RGB}{206,214,224}

\newcommand{\product}{ZitPit}
\newcommand{\paperbox}[1]{%
\vspace{0.4em}\noindent%
\fcolorbox{boxstroke}{boxfill}{%
\parbox{\dimexpr\columnwidth-2\fboxsep-2\fboxrule\relax}{#1}}%
\vspace{0.45em}}
\newcolumntype{L}[1]{>{\raggedright\arraybackslash}p{#1}}
\newcolumntype{Y}{>{\raggedright\arraybackslash}X}

\begin{document}

\twocolumn[
\begin{@twocolumnfalse}
\begin{center}
{\LARGE\bfseries \product: Consumer-Side Admission Control for Agentic Software Intake\par}
\vspace{0.15em}
{\large Turning First-Seen External Artifacts into Policy Events\par}
\vspace{0.28em}
{\normalsize Jepson Taylor \hspace{0.75em} Chris Brousseau \hspace{0.75em} Jordan Hildebrandt \hspace{0.75em} Kelli Quinn\par}
\vspace{0.04em}
{\small \texttt{jepson@veox.ai} \hspace{0.55em} \texttt{chris@veox.ai} \hspace{0.55em} \texttt{j@veox.ai} \hspace{0.55em} \texttt{kelli@veox.ai}\par}
\vspace{0.10em}
{\small VEOX Research Group\par}
{\small \url{https://github.com/jeppsontaylor/zitpit}\par}
\vspace{0.22em}
\end{center}

\begin{abstract}
AI IDEs and coding agents compress discovery, fetch, workspace open, installation, and execution into one low-observability loop. Existing defenses such as provenance frameworks, package and repository firewalls, runtime protection, and tool-approval prompts each cover part of that path, but they often leave the final consumer-side execution decision implicit. \product{} is a 100\% open-source Rust system that argues for a stricter boundary: first-seen external artifacts should become durable policy events before they gain execution rights on protected developer or CI hosts. The current public evidence is intentionally narrow and explicit. It includes repeated Git smart-HTTP intake measurements showing that approved artifacts can remain faster than unmanaged public fetch, plus implemented protected-session and governed-egress proof families. The broader contribution is architectural rather than universal-coverage-by-assertion: \product{} unifies artifact admission, repo-open state, capability-scoped execution, and durable policy records at the consumer execution boundary for agentic workflows.
\end{abstract}

\vspace{0.08em}
\noindent\textbf{Index Terms}---software supply chain security, AI agents, admission control, provenance, quarantine, governed execution, observability
\vspace{0.12em}
\begin{center}
\resizebox{0.94\textwidth}{!}{\includegraphics{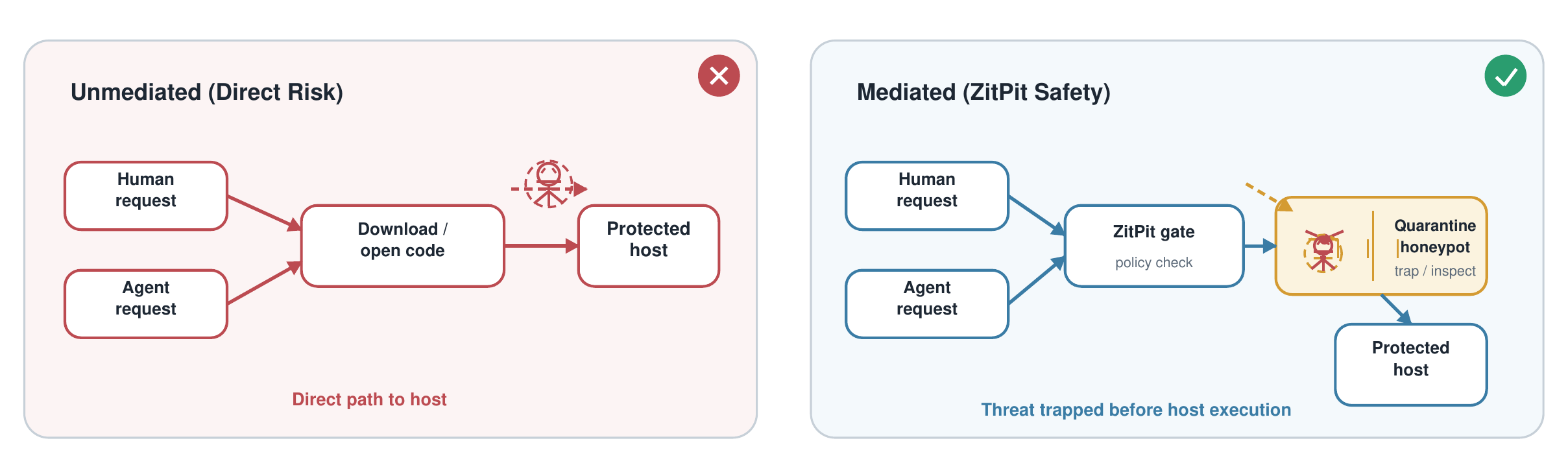}}

\vspace{0.1em}
\parbox{0.94\textwidth}{\footnotesize \textbf{Fig. 1.} Without mediation, human and agent requests can move newly encountered external software directly toward host execution; with \product{}, first-seen artifacts are held behind policy check, capability grant, and quarantine when required.}
\end{center}
\vspace{0.12em}
\end{@twocolumnfalse}
]

\section{Introduction}

Agentic development has changed the tempo of software intake. A developer can ask an AI IDE to ``set up this repo,'' and the tool may immediately clone a repository, open project memory files, attach MCP servers, evaluate workspace configuration, install dependencies, run build hooks, and invoke helper scripts before a human reaches a durable review checkpoint. The security problem is no longer only that a package might be malicious. It is that discovery, fetch, repo-open state, installation, and execution now collapse into one conversational loop with weak artifact-level observability.

This is visible in current tooling. Claude Code documentation covers npm installation, MCP configuration, hooks, project memory, devcontainers, and settings that shape tool use or execution surfaces \cite{anthropicGettingStarted2026,anthropicMcp2026,anthropicHooks2026,anthropicMemory2026,anthropicDevcontainer2026,anthropicSettings2026}. GitHub Actions guidance insists that third-party actions should be pinned to immutable SHAs because mutable references shift trust exactly where automation is strongest \cite{githubActionsHardening2026,githubReusableWorkflows2026}. VS Code Workspace Trust likewise exists because opening a folder can enable tasks, settings, extensions, and agents that influence execution \cite{vscodeWorkspaceTrust2026,vscodeCopilotSecurity2026}.

The resulting gap is consumer-side. Provenance systems can describe identity and build lineage. Repository and package firewalls can screen particular ecosystems. Runtime protection can watch after code begins executing. Command approval prompts can limit some tool uses. But in agentic workflows these controls still often stop short of one durable question: \emph{has this exact first-seen external artifact earned execution rights on this protected host, under this policy, in this context?} \cite{tufDocs2026,sigstoreDocs2026,intotoDocs2026,slsaDocs2026,sonatypeRepositoryFirewall2026,datadogSupplyChainFirewall2026,socketFirewall2026,stepSecurityHardenRunner2026}

\paperbox{\textbf{Thesis.} \product{} is a consumer-side software admission control layer for agentic development. First-seen external artifacts must earn execution rights under durable policy before they affect a protected host.}

The paper's contribution is not the claim that \product{} already closes every package manager, IDE, workflow engine, or runtime path. The contribution is narrower and more defensible: \product{} identifies a missing enforcement boundary for agentic development and shows preliminary public evidence that this boundary can be made visible, policy-scoped, and fast enough to be deployable.

\section{Why Admission Control Matters}

The importance of this boundary is architectural. Provenance only becomes operational when a consumer-side system uses it to decide whether execution rights will be granted. Repo-open state matters because opening a repository can change tool behavior before durable review. Capability-scoped verdicts matter because fetch, build, test, and host execution are different trust decisions. The safe path must be faster than unmanaged public fetch because mandatory controls that lose on latency are routinely bypassed.

\paperbox{\textbf{Novelty claim.} \product{} does not claim to invent application control, sandboxing, or provenance. Its novelty is the combination of artifact admission, repo-open state, capability-scoped execution, and durable policy events at the consumer execution boundary for agentic workflows.}

This matters especially for smaller organizations. Large enterprises can sometimes absorb fragmented controls through internal mirrors, release engineering, and custom policy teams. Smaller teams face the same machine-speed intake dynamics without the same guardrails. OpenSSF's guidance for AI code assistants emphasizes prompt injection, package confusion, and automation risk for exactly this reason \cite{openssfAICode2025}.

\section{Current Proof Boundary}

The strongest way to present \product{} is to separate current proof from future ambition. Table~\ref{tab:proofboundary} summarizes the current public evidence boundary that this paper relies on.

\begin{table*}[t]
\caption{Current public proof boundary}
\label{tab:proofboundary}
\centering
\scriptsize
\setlength{\tabcolsep}{4pt}
\begin{tabularx}{\textwidth}{@{}L{0.18\textwidth}L{0.27\textwidth}L{0.10\textwidth}Y@{}}
\toprule
\textbf{Surface} & \textbf{Current public evidence} & \textbf{Status} & \textbf{Supported claim} \\
\midrule
Git smart-HTTP intake & Repeated five-repository benchmark harness measuring web, disk-cache, and hot-cache latency \cite{zitpitBench2026} & Implemented & Approved immutable Git intake can stay faster than unmanaged public fetch \\
Brokered protected-session enforcement families & Docker demo + battle packs for blocked secret reads, direct egress tooling, persistence writes, publish abuse, destructive ops, and interpreter-evasion wrappers \cite{zitpitBenchmarkMatrix2026} & Implemented & Protected sessions can deny selected high-value command families before execution \\
Governed outbound DLP & Demo smoke proofs and battle packs for blocked sensitive uploads and archive scanning \cite{zitpitBenchmarkMatrix2026} & Implemented & Governed egress can block selected sensitive outbound data before transmission \\
Rust build-time execution & Battle harness coverage for \texttt{build.rs} style scenarios \cite{zitpitBenchmarkMatrix2026} & Partial & Build-time execution can be modeled as a separate capability boundary \\
GitHub Actions immutable-ref enforcement & Threat-model and scenario coverage for mutable refs and unsafe actions \cite{zitpitBenchmarkMatrix2026} & Partial & Workflow references should resolve to immutable identities before execution \\
npm / PyPI / raw installer mediation & Benchmark matrix and roadmap targets only \cite{zitpitBenchmarkMatrix2026} & Planned & Current paper does not claim ecosystem-complete package-manager enforcement \\
Repo-open enforcement depth & Threat model, policy model, and roadmap targets for \texttt{.mcp.json}, memory files, hooks, and devcontainers \cite{zitpitBenchmarkMatrix2026} & Planned & Repo-open state is part of the supply chain, but host-side closure is not yet fully proven \\
\bottomrule
\end{tabularx}
\end{table*}

\paperbox{\textbf{What \product{} does not claim.} \product{} is not a general agent-safety system. It does not prove that unknown software is benign, does not claim full ecosystem closure today, does not make trusted-publisher compromise harmless, and does not claim safety for unsupported or unmanaged paths.}

\section{Threat Model and Definitions}

\paperbox{\textbf{Definitions.} An \emph{artifact} is the external code object or repo-scoped execution bundle being mediated: a Git ref target, package tarball, wheel, crate source, workflow action, raw installer payload, or repo-open configuration surface. \emph{First-seen} means first observed by the ZitPit-mediated trust domain for a given immutable identity. A \emph{policy event} is the durable record created when a selector is resolved, evaluated, and granted or denied capability. A \emph{protected host} is a developer or CI environment whose execution rights are gated by \product{}.}

\subsection{Mandatory Mediation and Bypass Assumptions}

The current guarantee depends on mandatory mediation or transparent redirection for the paths under protection. This is the hardest part of the architecture and the place where skeptical reviewers will press first. Git submodules, partial clone follow-on fetches, Git LFS hydration, browser downloads, vendored tarballs, alternate registries, local copies, and direct unmanaged egress all weaken the guarantee if they escape policy visibility \cite{gitSubmodules2026,gitPartialClone2026,gitLfs2026,pipVcsSupport2026,cargoSourceReplacement2026}

\subsection{Transitive Closure and Delayed Resolution}

Top-level pinning is not enough when transitive dependencies, build backends, workflow reuse, or devcontainer features discover more artifacts later. npm Git dependencies can trigger lifecycle behavior; Python sdists can pull dynamic build requirements; Cargo build-time execution can extend beyond simple fetch; workflow references can expand through reusable workflows or mutable tags \cite{npmPackageJson2026,pipSecureInstalls2026,cargoBuildScripts2026,githubReusableWorkflows2026}

\subsection{Repo-Open and Workspace Execution Surfaces}

Repo-open state is in scope because opening a project can be execution-relevant. MCP definitions, hooks, memory files, startup tasks, devcontainers, and host-side initialization hooks can all shape behavior or launch follow-on actions before human review \cite{anthropicMcp2026,anthropicHooks2026,anthropicMemory2026,anthropicDevcontainer2026,devcontainerSpec2026,vscodeWorkspaceTrust2026}

\subsection{Trust Plane, Cache, and Control-Plane Compromise}

The control plane becomes part of the trusted computing base. Cache poisoning, policy-store compromise, signing-key compromise, parser weakness, and stale trust roots are all relevant risks. \product{} therefore treats resolved immutable identity, compatibility fingerprints, and any separately computed content digests as distinct inputs rather than collapsing them into one trust bit. Provenance, freshness, expiry, revocation, and operator-visible evidence remain separate obligations \cite{tufDocs2026,tufPaper2016,sigstoreDocs2026,slsaDocs2026}

\subsection{Operator Burden and Availability}

Admission control creates operational cost. Approval latency, break-glass frequency, policy drift, and availability all matter because a brittle control plane will be bypassed. The current paper therefore treats deployability as part of the security argument rather than as a UX-only concern.

\section{Architecture and Policy}

\subsection{Admission Stages}

\product{} organizes the control plane into four stages:

\begin{enumerate}
    \item \textbf{Acquire.} Resolve external requests to the strongest available immutable identity.
    \item \textbf{Build.} Separate build-time or install-time execution from simple acquisition.
    \item \textbf{Execute.} Grant capabilities rather than ambient trust to host execution surfaces.
    \item \textbf{Publish.} Optionally inspect outgoing release artifacts and workflow outputs.
\end{enumerate}

The current implementation is strongest on Git-path intake, protected-session mediation, and governed egress. Broader ecosystem adapters remain a public engineering agenda rather than hidden proof.

\subsection{Artifact Policy Events}

The admission decision is durable. Each artifact policy event records the requested selector, resolved immutable identity, provenance result, verdict, evidence pointer, context, and expiry or revocation state. That durable event is the missing join key for recall, audit, and later incident reconstruction.

\paperbox{\textbf{Policy-event schema.} selector; resolved immutable identity; provenance result; verdict; evidence pointer; context; expiry and revocation state. This is the shared contract between admission, evidence, and later recall.}

\paperbox{\textbf{Example event.}\\
selector = \texttt{acme/tool@\{pre-resolution\}}\\
resolved immutable identity = \texttt{f3c1...}\\
provenance result = \texttt{verified}; verdict = \texttt{RUN\_DEV}\\
evidence pointer = \nolinkurl{report://quarantine}\\
context = \nolinkurl{code_intake/protected_host}\\
expiry and revocation recorded at decision time.}

\subsection{Capability-Scoped Verdicts}

Binary allow or block semantics are too coarse for agentic workflows. \product{} uses capability-scoped verdicts so that fetching bytes, unpacking, building in quarantine, testing without secrets, and running on a protected host are separate decisions.

\begin{table}[t]
\caption{Capability-scoped verdicts}
\label{tab:verdicts}
\centering
\scriptsize
\setlength{\tabcolsep}{2pt}
\begin{tabularx}{\columnwidth}{@{}L{0.25\columnwidth}L{0.24\columnwidth}Y@{}}
\toprule
\textbf{Verdict} & \textbf{Typical use} & \textbf{Execution conditions} \\
\midrule
\texttt{FETCH\_ONLY} & First-seen dependency or repo bundle & Download allowed; protected-host execution denied \\
\texttt{UNPACK\_ONLY} & Archive expansion and static inspection & Unpack allowed; execution denied \\
\shortstack[l]{\texttt{BUILD\_NO\_}\\\texttt{NETWORK}} & sdist build, \texttt{build.rs}, installer analysis & Controlled lane only; no general egress \\
\texttt{TEST\_NO\_SECRETS} & Benign validation & Isolated test lane; no sensitive material \\
\texttt{RUN\_DEV} & Approved developer dependency & Protected-host execution under freshness policy \\
\texttt{RUN\_CI} & Approved CI dependency or action & CI execution under stronger identity and expiry checks \\
\texttt{BLOCKED} & Malicious, unsupported, or stale input & Denied pending recall, expiry, or operator action \\
\bottomrule
\end{tabularx}
\end{table}

\subsection{Protected-Session Enforcement Families}

The repository currently demonstrates protected-session enforcement families rather than universal host guarantees. The battle packs and Docker demo cover representative pre-execution denials for interpreter-evasion wrappers, secret and key reads, SSH-agent touch, browser token access, repo-open or config abuse, publish and deploy misuse, persistence writes, destructive operations, and selected recon or lateral-movement tooling \cite{zitpitBenchmarkMatrix2026}. These are important proof families, but they should be described as demonstrated brokered-session controls unless host-side mandatory enforcement is fully proven.

\subsection{Mirage Lab as Evidence, Not Oracle}

Mirage Lab is useful because it improves ordering, evidence production, and operator review. It is not a safety oracle. Dynamic analysis can miss delayed triggers, sandbox-aware behavior, context-dependent payloads, or follow-on stages. The strongest statement \product{} can make is still that unknown artifacts do not gain protected-host execution rights before policy evaluation.

\subsection{Provenance Consumption}

\product{} is designed to consume standards-backed signals rather than replace them: TUF-style freshness and anti-rollback semantics, Sigstore identity and transparency, in-toto attestations, SLSA provenance, GitHub artifact attestations, npm trusted publishing, PyPI trusted publishing, and checksum-backed package ecosystems such as Go modules \cite{tufDocs2026,tufPaper2016,sigstoreDocs2026,intotoDocs2026,intotoPaper2019,slsaDocs2026,githubArtifactAttestations2026,npmTrustedPublishing2026,pypiTrustedPublishing2026,pypiAttestations2026,goChecksum2026}

\section{Preliminary Evaluation}

The current evaluation is split into two proof obligations: deployability and coverage honesty.

\subsection{Benchmark Methodology}

The public timing harness now mirrors the actual upstream repository at the resolved immutable target before timing the approved path. The harness validates that the seeded managed mirror matches the claimed upstream HEAD, and it fails report generation if the managed response diverges from the expected immutable target. The mutable working outputs remain under \texttt{docs/benchmarks/latest.*}, but the paper cites a frozen benchmark snapshot artifact under \texttt{docs/benchmarks/snapshots/} so the evidence reference is not a moving ``latest'' pointer \cite{zitpitBench2026}.

\subsection{Deployability: Can the Safe Path Stay Fast?}

The public benchmark harness measures the Git smart-HTTP intake path rather than a full clone or cross-ecosystem install. It times repeated \texttt{git ls-remote} / \texttt{info/refs} style mediation across five public repositories: \texttt{git}, \texttt{go}, \texttt{node}, \texttt{cpython}, and \texttt{terraform}. Each repository is measured in three modes: direct upstream request (\texttt{web}), approved disk-cache hit (\texttt{cache}), and approved hot-cache hit (\texttt{hot-cache}) \cite{zitpitBench2026}.

\begin{figure*}[t]
\centering
\includegraphics[width=0.98\textwidth]{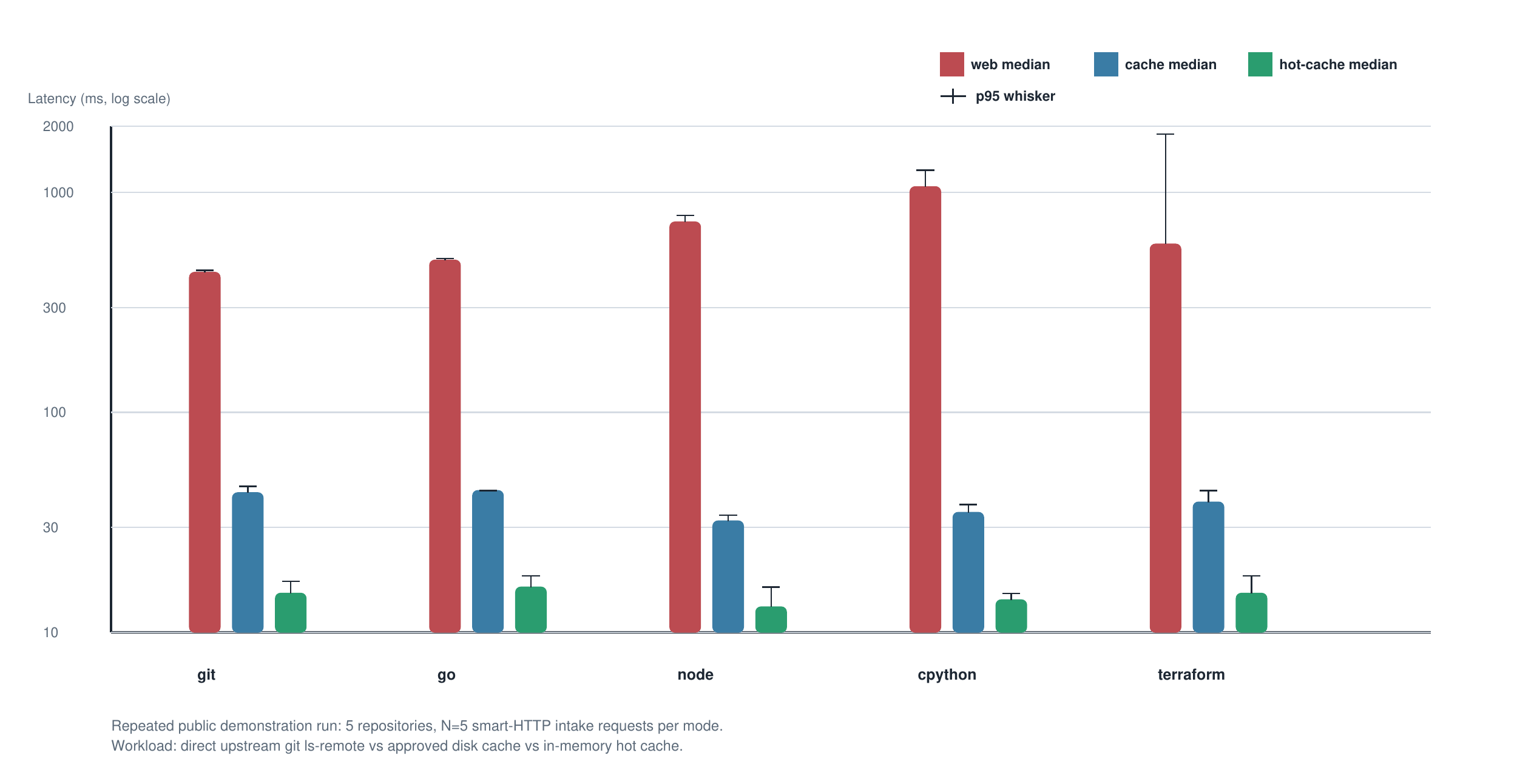}
\caption{Preliminary deployability result for the Git smart-HTTP intake path. Repository-level web medians ranged from 433--1062~ms, cache medians from 32--44~ms, and hot-cache medians from 13--16~ms across \(N=5\) samples per repository \cite{zitpitBench2026}. The significance is not universal Git closure; it is evidence that approved immutable intake can be materially faster than unmanaged public fetch.}
\label{fig:speed}
\end{figure*}

This result is deliberately modest. It does not prove full clone closure, submodule closure, LFS closure, or package-manager-native mediation. It does show something operationally important: if admission control is to survive contact with real development workflows, approved artifacts cannot be slower than the public network by default.

\subsection{Coverage Honesty: Which Attack Families Are Publicly Supported?}

\begin{table*}[t]
\caption{Attack-family coverage and current proof status}
\label{tab:coverage}
\centering
\scriptsize
\setlength{\tabcolsep}{3pt}
\begin{tabular}{@{}L{0.13\textwidth}L{0.14\textwidth}L{0.16\textwidth}L{0.10\textwidth}L{0.18\textwidth}L{0.21\textwidth}@{}}
\toprule
\textbf{Surface} & \textbf{First execution boundary} & \textbf{Control point} & \textbf{Status} & \textbf{Evidence source} & \textbf{Residual risk} \\
\midrule
Git smart-HTTP intake & checkout or follow-on fetch & mediated intake and immutable ref binding & Implemented & repeated public benchmark harness \cite{zitpitBench2026} & submodules, LFS, and follow-on fetch are not yet fully closed \\
Protected-session command families & brokered shell before process start & session broker policy and battle packs & Implemented & protected-session battle packs \cite{zitpitBenchmarkMatrix2026} & host-side mandatory enforcement remains future work \\
Governed outbound DLP & proxy before upstream routing & egress broker and DLP classifiers & Implemented & smoke proofs and egress battle packs \cite{zitpitBenchmarkMatrix2026} & raw sockets and unmanaged egress remain outside current proof \\
Rust build-time execution & compile-time build script & controlled build lane & Partial & battle harnesses \cite{zitpitBenchmarkMatrix2026} & broader Cargo-native mediation remains future work \\
GitHub Actions refs and unsafe actions & workflow runner & immutable ref policy and scenario coverage & Partial & workflow scenarios \cite{zitpitBenchmarkMatrix2026} & reusable workflows and full graph closure remain incomplete \\
npm / PyPI / raw installers & install, build, or lifecycle hook & package-manager-native mediation & Planned & benchmark matrix + roadmap \cite{zitpitBenchmarkMatrix2026} & native package-manager closure not yet public proof \\
Repo-open surfaces & workspace open or tool startup & workspace policy + agent policy hooks & Planned & threat model + roadmap \cite{zitpitBenchmarkMatrix2026} & enforcement depth varies by IDE and runtime \\
\bottomrule
\end{tabular}
\end{table*}

The coverage matrix is part of the argument, not an embarrassment to hide. A credible paper should make unsupported or partial paths visible rather than imply full closure through architectural rhetoric.

\balance
\section{Related Work}

\product{} overlaps with several adjacent control families, but it is not reducible to any one of them.

\textbf{Provenance and attestations.} TUF, Sigstore, in-toto, SLSA, GitHub artifact attestations, npm trusted publishing, and PyPI trusted publishing improve identity, freshness, transparency, and build-lineage statements \cite{tufDocs2026,tufPaper2016,sigstoreDocs2026,intotoDocs2026,intotoPaper2019,slsaDocs2026,githubArtifactAttestations2026,npmTrustedPublishing2026,pypiTrustedPublishing2026,pypiAttestations2026}. \product{} aims to consume those signals at the moment execution rights are granted.

\textbf{Workspace trust and application control.} VS Code Workspace Trust, Windows application control, and platform notarization traditions already recognize that opening untrusted content or launching unsigned code can change host risk posture \cite{vscodeWorkspaceTrust2026,windowsAppControl2026,appleGatekeeper2026}. \product{} differs by centering the consumer-side admission event for heterogeneous developer artifacts and repo-open bundles in agentic workflows.

\textbf{Hermetic and reproducible builds.} Bazel-style hermeticity, checksum-backed ecosystems such as Go modules, and package-manager-native integrity controls show that locality, reproducibility, and strong identity can improve both speed and trust \cite{bazelHermeticity2026,goChecksum2026,pipSecureInstalls2026,cargoSourceReplacement2026}. \product{} extends that intuition to the broader moment where external software, repo-open state, and workflow references receive host rights.

\textbf{Repository and package firewalls.} Sonatype Repository Firewall, Datadog Supply-Chain Firewall, and Socket Firewall provide important ingress screening and quarantine functions \cite{sonatypeRepositoryFirewall2026,datadogSupplyChainFirewall2026,socketFirewall2026}. \product{} shares that concern, but pushes toward a unified admission record across intake, repo-open state, execution, and recall.

\textbf{Runtime protection.} Tools such as StepSecurity Harden-Runner observe or constrain execution after code begins running \cite{stepSecurityHardenRunner2026}. Those controls remain valuable, but \product{} argues that the first durable consumer-side decision should occur earlier.

\textbf{Behavioral analysis corpora.} OpenSSF Package Analysis and malicious-package corpora help quantify file access, command execution, and network behavior across suspicious packages \cite{openssfPackageAnalysis2026,openssfMaliciousPackages2026}. \product{} treats that style of analysis as evidence input and benchmark material, not as a replacement for admission control.

\section{Implications}

The current paper makes a narrow empirical claim and a broader architectural claim.

Empirically, the strongest public evidence today is that some important slices of intake, protected-session mediation, and governed egress can be made explicit and auditable, while approved immutable intake can remain faster than unmanaged public fetch.

Architecturally, this pattern could become a standard execution boundary for agentic environments. The next frontier is broader admission of external influence, not just packages: repo-open state, workflow references, tool manifests, and other machine-consumable context increasingly behave like supply-chain input. If admission systems become part of mainstream development infrastructure, open governance matters because these systems can centralize power if their policy memory, recall logic, and evidence formats are opaque or non-portable.

\section{Limitations and Future Work}

\textbf{Coverage remains incomplete.} Current public proof is strongest for Git smart-HTTP intake, protected-session command families, and governed egress. Package-manager-native closure, repo-open enforcement depth, raw installer capture, and full workflow graph closure remain future work.

\textbf{Mandatory mediation remains hard.} Unsupported or unmanaged paths should be called unsupported, not quietly treated as secure enough.

\textbf{Trusted publishers can still ship bad code.} Admission control narrows the moment when rights are granted; it does not make publisher compromise harmless.

\textbf{Mirage Lab is not a verifier.} Dynamic analysis improves evidence and ordering but cannot prove benign behavior.

\textbf{Operational cost matters.} Break-glass controls, approval burden, and control-plane availability are part of the security story because systems that are too brittle are bypassed.

The near-term engineering agenda is concrete: broader package-manager-native mediation; stronger follow-on Git closure for submodules, LFS, and delayed fetches; repo-open benchmark families; stronger provenance consumption with expiry, revocation, and recall; and public benchmark suites that distinguish implemented proof from roadmap ambition.

\section{Conclusion}

\product{} starts from a simple observation about agentic development: the interval between discovering external software and granting it local authority is collapsing toward zero. That changes the practical trust problem. The important question is no longer only whether a package, workflow, or repo-open bundle exists on the internet. The important question is whether it has earned execution rights in a protected environment under durable policy.

The present-tense claims in this paper are intentionally narrow. First-seen external artifacts should become policy events before they gain protected-host execution rights. Approved immutable intake can stay on a faster path than unmanaged public fetch. Current public proof is strongest for selected intake, protected-session, and governed-egress surfaces, not for universal ecosystem closure.

The larger significance is still worth stating carefully. \product{} identifies a consumer-side admission boundary that adjacent fields have been approaching from different directions: provenance, package firewalls, workspace trust, runtime protection, and agent governance. If the agent era needs a durable place where outside software earns local authority, that boundary is a strong candidate.

\bibliographystyle{IEEEtran}
\bibliography{references}

\end{document}